\documentclass[twocolumn,showpacs,pre]{revtex4}
\usepackage{amsmath}
\usepackage{graphicx}
\usepackage{bm}
\usepackage{color}

\begin{document}
\title{Anisotropic spatially heterogeneous dynamics on the $\alpha$ and $\beta$ 
relaxation time scales studied via a four-point correlation function}

\author{Elijah Flenner and Grzegorz Szamel}

\affiliation{Department of Chemistry, Colorado State University, Fort Collins, CO 80523}
\date{\today}

\begin{abstract}
We examine the anisotropy of a four-point correlation function $G_4(\vec{k},\vec{r};t)$
and it's associated structure factor $S_4(\vec{k},\vec{q};t)$ calculated using
Brownian Dynamics computer simulations of a
model glass forming system. These correlation functions measure the spatial correlations
of the relaxation of different particles, and we examine the time and temperature 
dependence of the anisotropy. We find that the anisotropy is strongest at nearest
neighbor distances at time scales corresponding to the peak of the 
non-Gaussian parameter $\alpha_2(t) = 3 \langle \delta r^4(t) \rangle/[ 5 \langle \delta r^2(t) \rangle^2] - 1$, 
but is still pronounced around the $\alpha$
relaxation time. We find that the structure factor $S_4(\vec{k},\vec{q};t)$ is anisotropic
even for the smallest wave vector accessible in our simulation suggesting that our
system (and other systems commonly used in computer simulations) may be too small to
extract the $\vec{q} \to 0$ limit of the structure factor. 
We find that the determination of a dynamic correlation length from $S_4(\vec{k},\vec{q};t)$
is influenced by the anisotropy. We extract an effective anisotropic dynamic correlation length from 
the small $q$ behavior of $S_4(\vec{k},\vec{q};t)$.
\end{abstract}


\maketitle


\section{Introduction}
\label{sec:intro}
It is now generally accepted that upon approaching the glass
transition, the liquid's dynamics are becoming increasingly heterogeneous 
\cite{Ediger2000,Richert2002r,Andersen2005}. 
However, the details of the spatial and temporal characteristics 
of dynamic heterogeneities are still being debated. In particular, 
the connection between heterogeneous dynamics and a growing dynamic correlation length
has been the topic of many simulations 
\cite{Berthier2004,Berthier2007,Berthier2007p2,Biroli2006,Donati1998,Lacevic2003,Lacevic2002,Toninelli2005} 
and a few experimental studies
\cite{Berthier2006,Dalle-Ferrier2007,Lechenault2008}. Four-point correlation
functions have been introduced to facilitate the quantitative description of heterogeneous dynamics.
The analysis of the spatial decay of these correlation 
functions was used to extract a dynamic correlation length. Recently, the mode-coupling theory has been 
extended and a theoretical treatment of four-point correlation functions is starting to emerge 
\cite{Biroli2004,Berthier2006,Berthier2007p2,Berthier2006p3,Szamel2008,Iwata2007}. 
However, in most simulation studies and in some theoretical treatments these 
four-point correlation functions have been assumed to be isotropic or they are isotropic by design.

Researchers have noticed anisotropy in the correlated motion of particles on the $\beta$ 
relaxation time scale, 
and recently this anisotropic motion has also been reported on the
$\alpha$ relaxation time scale \cite{Flenner2007}. Doliwa and Heuer 
\cite{Doliwa2000} reported anisotropic correlated 
motion in a hard sphere system on the $\beta$ relaxation
time scale. Anisotropic motion has also been extensively studied by Donati \textit{et al.}\
and Gebremichael \textit{et al.}\ \cite{Donati1998,Gebremichael2004} who described 
the motion of "mobile" particles as "string-like", with mobile particles following 
each other in one dimensional "strings". Weeks \textit{et al.}  \cite{Weeks2002} have reported anisotropic dynamics
associated with the break down of the "cage" surrounding a particle. They found that the
correlations of the particle's displacements depends on the initial separation of the particles. 
While particles that start at a separation corresponding the the first peak of the pair
correlation function are most likely to move in the same direction, particles that start 
at a separation corresponding to the first minimum are more likely to initially move in opposite 
directions. 

In view of the experimental and simulational evidence for anisotropic correlations of 
particle's displacements, it should not be a surprise that four-point correlation functions designed
to study these dynamics can also be anisotropic. 
However, this anisotropy is normally studied for times less than the $\alpha$ relaxation time,  
thus it is uncertain if understanding this 
anisotropy is important for the structural relaxation of the  liquid. Previously \cite{Flenner2007}
we reported on a four-point correlation function that is anisotropic on the $\alpha$ relaxation
time scale as well as the $\beta$ relaxation time scale for a model glass forming liquid. Since the
spatial decay of this correlation function can be used to determine a dynamic length scale, the 
anisotropy introduces a complication in determining this length scale. 

In this paper we expand on previous work \cite{Flenner2007}. 
After describing the simulation in Sec.~\ref{sec:sim},
we explore the anisotropic correlated dynamics by examining a four-point correlation
function $G_4(\vec{k},\vec{r};t)$, Sec.~\ref{sec:fourpoint}, and the associated structure factor 
$S_4(\vec{k},\vec{q};t)$, Sec.~\ref{sec:sfactor}. 
We examine the anisotropy at around nearest neighbor 
distances, which corresponds to local rearrangement of particles and its cage, 
and at large distances. We examine how the anisotropy influences the determination of 
a growing length scale accompanying the glass transition, and determine an effective anisotropic
correlation length.  We finish with a discussion
of the results in Sec.~\ref{sec:conclusions}.
\section{Simulation}
\label{sec:sim}
We performed Brownian dynamics simulations of an 80:20 binary mixture of 1000 particles introduced by 
Kob and Andersen \cite{Kob1995,Kob1995II}. The interaction potential is 
$V_{\alpha \beta}(r) = 4 \epsilon_{\alpha \beta}[ (\sigma_{\alpha \beta}/r)^{12} - (\sigma_{\alpha \beta}/r)^6 ]$ 
where $\alpha$, $\beta \in$ \{A,B\}, $\epsilon_{AA} =$ 1.0, $\epsilon_{AB} =$ 1.5, 
$\epsilon_{BB} =$ 0.5, $\sigma_{AA}$ = 1.0, $\sigma_{AB} =$ 0.8, and $\sigma_{BB} =$ 0.88 and
the interaction potential is cut at 2.5 $\sigma_{\alpha \beta}$. Periodic boundary 
conditions were used with a box length of 9.4 $\sigma_{AA}$. 
The equation of motion for the position of particle $i$ is
\begin{equation}
\dot{\vec{r}}_i = \frac{1}{\xi_0} \vec{F}_i(t) + \vec{\eta}_i(t),
\label{eq:motion}
\end{equation}
where $\xi_0 = 1.0$ is the friction coefficient of an isolated particle and the force acting on
a particle $i$ is
\begin{equation}
\vec{F}_i = -\nabla_i \sum_{n \ne i} V_{\alpha \beta} (|\vec{r}_i - \vec{r}_n|)
\end{equation}
with $\nabla_i$ being the gradient operator with respect to $\vec{r}_i$. 
The random force $\vec{\eta}(t)$ satisfies the fluctuation dissipation relation
\begin{equation}
\langle \vec{\eta}_i(t) \vec{\eta}_j(t^\prime) \rangle = 2 D_0 \delta_{ij} \mathbf{1},
\end{equation}
where $D_0 = k_B T/\xi_0$, $k_B$ is Boltzmann's constant, and $\mathbf{1}$ is the unit tensor.
The results are presented in terms of reduced units with $\sigma_{AA}$, $\epsilon_{AA}/k_B$,
and $\sigma_{AA}^2 \xi_0/\epsilon_{AA}$ being the units of length, energy, and time, 
respectively. Since the equation of motion allows for diffusion of the center of mass, all results 
are presented relative to the center of mass.

We present results for temperatures $T = 0.45$, 0.47, 0.5, 0.55, 0.6, 0.8, 0.9, and 1.0. 
The onset of supercooling is around $T=1.0$ and we use $T_{c} = 0.435$
as the mode coupling temperature. As a means to expand the temperature scale, we 
will plot some quantities versus $\epsilon = (T-T_{c})/T_{c}$. 
The equation of motion was integrated using a Heun algorithm with a small time step of 
$5\times 10^{-5}$. We ran an equilibration run that was at least half as long as a production run,
and four production runs at each temperature. The results are an average over the production runs.
We present results only for the larger and more abundant $A$ particles. 
We define the $\alpha$ relaxation time $\tau_\alpha$ 
as through relation $F_s(\vec{k};\tau_\alpha) = e^{-1}$ for a wave vector around
the first peak of the partial static structure factor for the $A$ particles, which corresponds to $|\vec{k}| = 7.25$.

\section{Four-point correlation function $G_4(\vec{k},\vec{r};t)$}
\label{sec:fourpoint}
\subsection{Definition and connection with overlap correlations}
We study a four-point correlation function that measures the spatial and 
temporal correlations between the relaxation 
of different particles. Consider the function
\begin{equation}
\hat{F}_n(\vec{k};t) = e^{-i\vec{k} \cdot [\vec{r}_n(t) - \vec{r}_n(0)]},
\end{equation}
where $\vec{r}_n(t)$ is the position of particle $n$ at a time $t$. 
The ensemble average of $\hat{F}_n(\vec{k};t)$ is the self-intermediate scattering function $F_s(k;t)$,
thus we will term $\hat{F}_n(\vec{k};t)$ the microscopic self-intermediate scattering function. 
The four-point correlation function
\begin{equation}
G_4(\vec{k},\vec{r};t) = \frac{V}{N^2} \sum_{n\ne m}\langle \hat{F}_n(\vec{k};t) \hat{F}_m(-\vec{k};t) 
\delta[ \vec{r} - \vec{r}_{nm}(0)] \rangle
\label{eq:gfour}  
\end{equation}
measures the correlations between the microscopic self-intermediate scattering function at time $t$, pertaining to 
particles that are separated 
by a vector $\vec{r}$ at the initial time. In Eq.~\eqref{eq:gfour} $\vec{r}_{nm} = \vec{r}_n - \vec{r}_m$, 
$V$ is the volume, and $N$ is the number of particles. Notice that $G_4(\vec{k},\vec{r};0) = g(r)$ 
where $g(r)$ is the pair correlation function. In this work we choose $|\vec{k}\,|$ to have the same value
as the one that determines the $\alpha$ relaxation time, \textit{i.e.} $|\vec{k}\,|$ is 
located around the first peak of the partial static structure factor for the $A$ particles, $|\vec{k}\,|  = 7.25$.

It should be noted that $G_4(\vec{k},\vec{r};t)$ is, in general, complex. Its real and imaginary
parts can be written in the following form 
\begin{eqnarray}
\lefteqn{Re[G_4(\vec{k},\vec{r};t)] = } \\ \nonumber && 
\frac{V}{N^2} \sum_{n\ne m}\left< \cos\{ \vec{k}\cdot[\vec{r}_{nm}(t) - \vec{r}_{nm}(0)] \}
\delta[ \vec{r} - \vec{r}_{nm}(0)] \right>
\label{eq:Rgfour}  
\end{eqnarray}
\begin{eqnarray}
\lefteqn{Im[G_4(\vec{k},\vec{r};t)] = } \\ \nonumber && 
-\frac{V}{N^2} \sum_{n\ne m}\left< \sin\{ \vec{k}\cdot[\vec{r}_{nm}(t) - \vec{r}_{nm}(0)] \}
\delta[ \vec{r} - \vec{r}_{nm}(0)] \right>
\label{eq:Igfour}  
\end{eqnarray}
Eqs. (\ref{eq:Rgfour}-\ref{eq:Igfour}) 
show that particles which are getting closer together or farther apart along the direction of 
vector $\vec{k}$ (\textit{i.e.} are moving in the opposite direction or in the same direction along $\vec{k}$)
make the same contribution to the real part of $G_4(\vec{k},\vec{r};t)$ but opposite contributions
to its imaginary part. In particular, particles moving farther apart along the direction of 
vector $\vec{k}$ make a negative contribution to the imaginary part of $G_4(\vec{k},\vec{r};t)$.

In several other simulational and experimental studies 
\cite{Lacevic2003,Lacevic2002,Bagchi2008,Abete2008}
four-point correlation functions involving single-particle overlaps rather than 
the microscopic self-intermediate scattering functions were investigated. For example, 
Lacevic \textit{et.\ al} \cite{Lacevic2003} used the following function \cite{Laceviccom}
\begin{equation}
g_4^{ol}(r;t) = \frac{V}{N^2} \sum_{n\ne m}\langle w_n(a;t) w_m(a;t) 
\delta[ \vec{r} - \vec{r}_{nm}(0)] \rangle,
\label{eq:Lgfour}  
\end{equation} 
where $w_n(a;t)$ is the overlap function pertaining to particle $n$,
\begin{equation}
w_n(a;t) = \theta(a-|\vec{r}_n(t) - \vec{r}_n(0)|).
\end{equation}
We would like to point out that $g_4(r;t)$ can be expressed in terms of functions which are
generalizations of our $G_4(\vec{k},\vec{r};t)$,
\begin{equation}
g_4^{ol}(r;t) = \int \frac{d\vec{k}_1 d\vec{k}_2}{(2\pi)^6} f(k_1;a) f(k_2;a) \mathcal{G}_4(\vec{k}_1,\vec{k}_2,\vec{r};t),
\end{equation}
where $\mathcal{G}_4(\vec{k}_1,\vec{k}_2,\vec{r};t)$ is defined as the correlation function
of the microscopic self-intermediate scattering function at time $t$
and calculated for different wave vectors,
\begin{equation}
\mathcal{G}_4(\vec{k}_1,\vec{k}_2,\vec{r};t) = \frac{V}{N^2} \sum_{n\ne m}\langle \hat{F}_n(\vec{k}_1;t) \hat{F}_m(\vec{k}_2;t) 
\delta[ \vec{r} - \vec{r}_{nm}(0)] \rangle,
\label{eq:newgfour}  
\end{equation}
and $f(k;a) = 4\pi a^2 j_1(ka)/k$ with $j_1$ denoting a spherical Bessel function of the first kind.

The present work is mostly concerned with the anisotropic nature of dynamic heterogeneities,
which can be monitored using the four-point correlation function 
given by Eq.~(\ref{eq:gfour}). In this context we would like to emphasize that
in principle the more general function (\ref{eq:newgfour}) is also anisotropic. However, any trace of this
anisotropy is lost after the integration over wave vectors $\vec{k}_1$ and $\vec{k}_2$ and thus
the overlap correlation function (\ref{eq:Lgfour}) is, by construction, isotropic.

\subsection{Anisotropy of $G_4(\vec{k},\vec{r};t)$}

\begin{figure}
\includegraphics[width=3.2in]{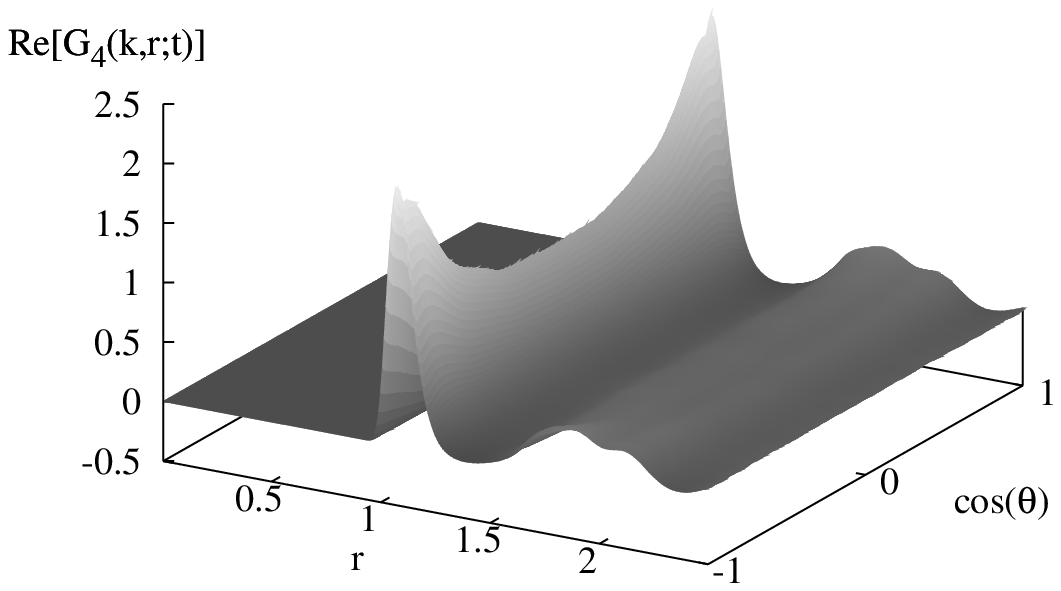}\\
\includegraphics[width=3.2in]{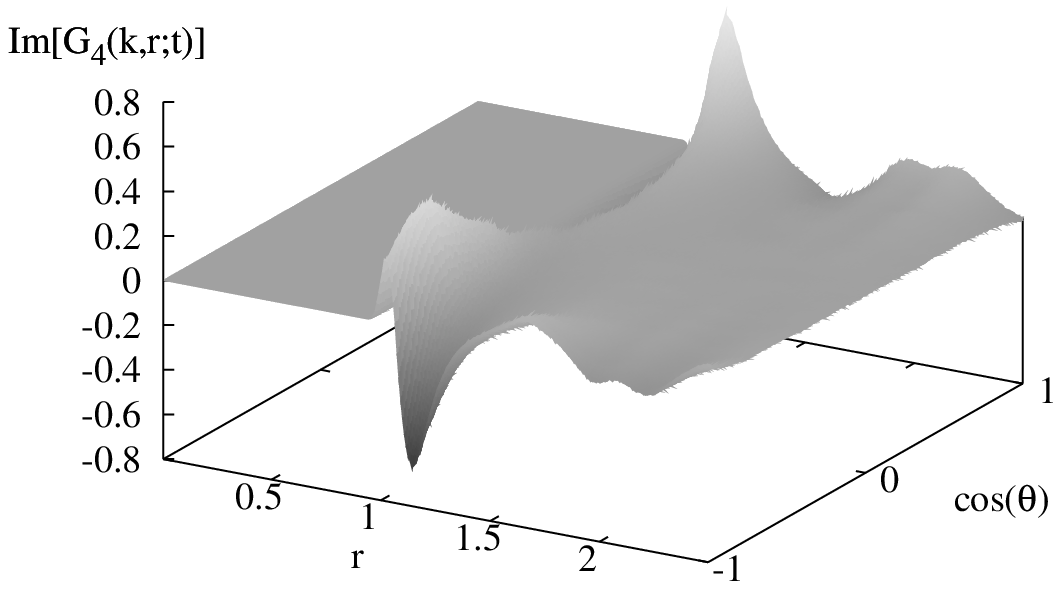}
\caption{The real part of the correlation function $G_4(\vec{k},\vec{r};\tau_\alpha)$ (upper figure) 
and the imaginary part of $G_4(\vec{k},\vec{r};\tau_\alpha)$ (lower figure) for T=0.45 
calculated at the $\alpha$ relaxation time.\label{fig:gfourtheta}}
\end{figure}

Since the functions $\hat{F}_n(\vec{k};t)$ are sensitive to displacements 
of particles along the direction of $\vec{k}$, then $G_4(\vec{k},\vec{r};t)$ measures 
interparticle correlations weighted by the displacements along the vector $\vec{k}$. 
Particles which move in the direction perpendicular to $\vec{k}$ make a contribution
to $G_4(\vec{k},\vec{r};t)$ which is the same as their contribution to the pair
correlation function $g(r)$.  
We notice that for $t>0$  four-point function $G_4(\vec{k},\vec{r};t)$ is not isotropic, but depends on the 
angle $\theta$ between $\vec{r}$ and $\vec{k}$. Shown in the upper figure 
in Fig.~\ref{fig:gfourtheta} is the real part $G_4(\vec{k},\vec{r};t)$
for $T=0.45$ calculated at $t = \tau_\alpha$, and the lower figure shows the imaginary part. The maximum value
of the real part of $G_4(\vec{k},\vec{r};\tau_\alpha)$ 
occurs for values of $\cos(\theta)$ corresponding to $\theta = 0^\circ$ and $\theta = 180^\circ$, which 
shows that the correlations are most pronounced for $\vec{r}$ parallel and antiparallel to $\vec{k}$. 
Thus, the correlations of the microscopic relaxation function is anisotropic on the 
the $\alpha$ relaxation time scale and the correlations are the strongest when neighboring particles
move in the same or in opposite directions. 

To examine these anisotropic correlations at length
scales around nearest neighbor distances, we expand $G_4(\vec{k},\vec{r};t)$ into the Legendre polynomials
\begin{equation}
G_4(\vec{k},\vec{r};t) = \sum_n L_n(k,r;t) P_n(\hat{\vec{k}} \cdot \hat{\vec{r}} ),
\end{equation}
where $P_n$ is the $n$th Legendre Polynomial, $\hat{\vec{k}} = \vec{k}/k, \hat{\vec{r}} = \vec{r}/r$,
and
\begin{equation}
L_n(k,r;t) = \frac{2n + 1}{4 \pi} \int G_4(\vec{k},\vec{r};t) P_n(\hat{\vec{k}} \cdot \hat{\vec{r}} ) 
\mbox{d}\hat{\vec{r}}.
\label{eq:expandcoeff}
\end{equation}
If $G_4(\vec{k},\vec{r};t)$ does not depend on the angle between $\vec{k}$ and $\vec{r}$, then 
$L_n(k,r;t)$ is zero for all $n$ not equal to zero. Since there are nonzero real and imaginary 
parts to $G_4(\vec{k},\vec{r};t)$ for $t > 0$, then there are nonzero real and imaginary parts to
$L_n(k,r;t)$. By symmetry, the imaginary part is zero for even $n$, and the real part is zero for 
odd $n$.

\begin{figure}
\includegraphics[width=5.8in,bb=0 0 792 612]{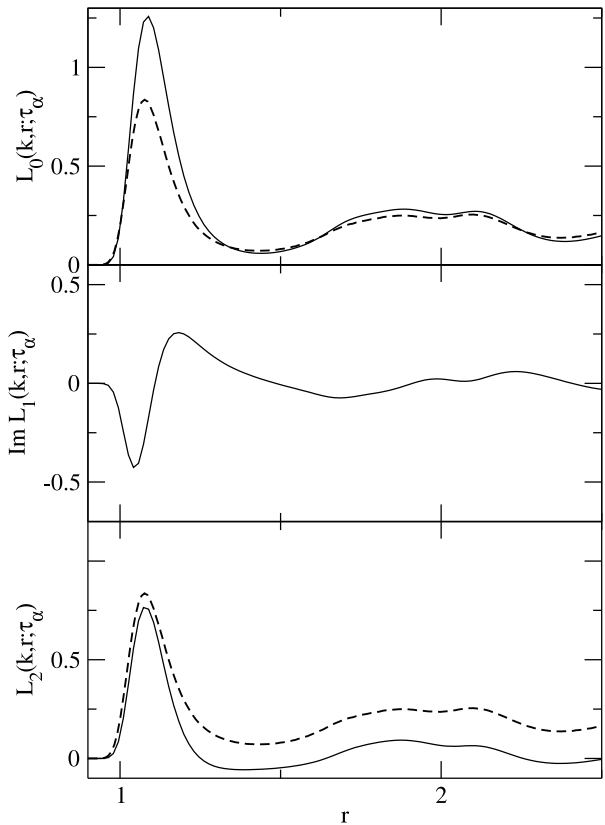}
\caption{The real part of $L_n(k,r;\tau_\alpha)$ for $n=0$ and 2, and the 
imaginary for $n=1$ for T=0.45 calculated at the $\alpha$ relaxation time. The dashed line in the figures
is $g(r)e^{-2}$ where $g(r)$ is the pair correlation function.\label{fig:Ln}}
\end{figure}

Shown in Fig.~\ref{fig:Ln} is the real part $L_n(k,r;\tau_\alpha)$ for $n = 0$ and 2, and the 
imaginary part for $n = 1$ at the alpha relaxation time $\tau_\alpha$ for $T=0.45$. There is a peak
in $L_2(k,r;\tau_\alpha)$ and $L_0(k,r;\tau_\alpha)$ around the first peak of the 
pair correlation function $g(r)$. The dashed lines in the figure are $g(r) e^{-2}$.
Note that due to our definition of the $\alpha$ relaxation time 
$e^{-2}$ is the asymptotic limit of the isotropic component $L_0$ at $t=\tau_{\alpha}$,  
$\lim_{r \rightarrow \infty} L_0(k,r;\tau_\alpha) = F_s^2(k,\tau_\alpha) = e^{-2}$. 
The positive peak in $L_2(k,r;\tau_\alpha)$ indicates that particles 
that are initially separated by a distance corresponding to the first peak of 
$g(r)$ have a tendency to move in the same direction or in opposite directions, while the values close to zero around 
the first minimum of the static structure factor can result from motion which is perpendicular 
to the initial separation vector. The spatial variation
of the correlated motion on these length scales has been reported previously in colloidal 
suspensions \cite{Weeks2002} and is related to the break up of the cage surrounding a particle. 

\begin{figure}
\includegraphics[width=3.2in,bb=0 0 792 612]{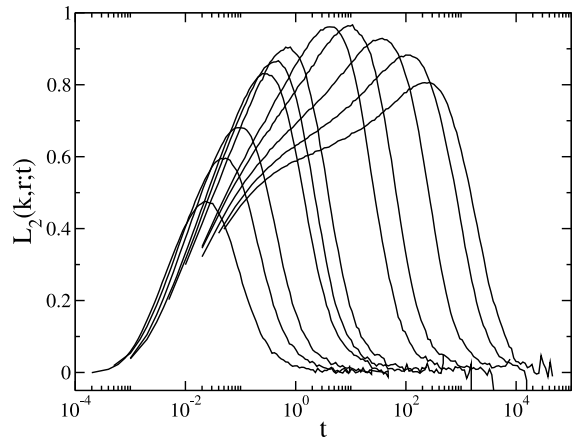}
\caption{The time dependence of the the first peak of $L_2(k,r;\tau_\alpha)$ 
for $T=1.0$, 0.9, 0.8, 0.6, 0.55, 0.5, 0.47, 0.45, shown from left to right.\label{fig:Ltime}}
\end{figure}

The variation of the imaginary part of $L_1(k,r;\tau_\alpha)$ indicates that particles closer 
than the first peak of $g(r)$ are more likely to move apart, while particles at a distance
greater than this peak are more likely to move closer together. In general, negative 
values of $L_1(k,r;t)$ indicates that particles move farther apart while positive values
indicate that particles move closer together. 

\begin{figure}
\includegraphics[width=3.2in,bb=0 0 792 612]{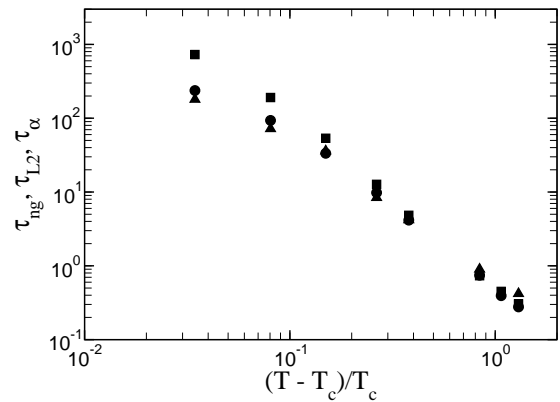}
\caption{The time at which the first peak of $L_2(k,r;t)$ reaches its maximum value, $\tau_{L2}$, 
(squares) compared to the $\alpha$ relaxation time, $\tau_{\alpha}$, (circles) and the peak time
of the standard non-gaussian parameter, $\tau_{ng}$, (triangles).\label{fig:Lpeak}}
\end{figure}

To look at the time dependence of the anisotropy, we calculated the height of the first
peak of $L_2(k,r;t)$ as a function of time, which is shown in
Fig.~\ref{fig:Ltime} for $T=1.0$, 0.9, 0.8, 0.6, 0.55, 0.5, 0.47 and 0.45. The peak height
starts at zero since the liquid is isotropic, then increases, reaches a maximum, 
and finally decreases to zero at long times. The height of the peak, $\tau_{L2}$, 
is around the $\alpha$ relaxation time for
high temperatures, Fig.~\ref{fig:Lpeak}, but its position increases slower with decreasing
temperature than the $\alpha$  relaxation time and approximately follows the
temperature dependence of the time corresponding to the peak position of the standard non-Gaussian 
parameter $\alpha_2(t) = 3 \langle \delta r^4(t) \rangle / [5 \langle \delta r^2(t) \rangle^2 ] -1$, 
$\tau_{ng}$ (triangles in Fig.~\ref{fig:Lpeak}). Furthermore, the maximum value
does not monotonically increase with a decrease in the temperature, but rather reaches a maximum
around $T=0.55$, then begins to decrease with decreasing temperature. Thus the anisotropy
around nearest neighbor distances initially increases upon
supercooling the liquid, but reaches a maximum and begins to slowly decrease when the 
liquid is cooled further.  It is not known if the peak height continues to decrease or saturates at 
low temperatures.

\section{Four-point structure factor $S_4(\vec{k},\vec{q};t)$}
\label{sec:sfactor}
\subsection{Anisotropy of $S_4(\vec{k},\vec{q};t)$}
To investigate the correlations between microscopic self-intermediate scattering functions 
at larger distances, we examined the the structure factor
corresponding to $G_4(\vec{k},\vec{r};t)$,
\begin{equation}
S_4(\vec{k},\vec{q};t) = 1 + \frac{N}{V} H_4(\vec{k},\vec{q};t)
\end{equation}\
where $H_4(\vec{k},\vec{q};t)$ is the Fourier transform of $G_4(\vec{k},\vec{r};t) - F_s^2(k;t)$.
For $\vec{q} \ne 0$
\begin{equation}
\label{eq:sfactor}
S_4(\vec{k},\vec{q};t) = \frac{1}{N} \sum_{n,m} \langle \hat{F}_n(\vec{k};t) \hat{F}_m(-\vec{k};t) 
e^{-i \vec{q} \cdot \vec{r}_{nm}(0)} \rangle.
\end{equation}
Again, we fix $|\vec{k}|$ to be around the position of the first peak of the static
structure factor for the $A$ particles, $|\vec{k}|\, = 7.25$.

\begin{figure}
\includegraphics[width=3.2in,bb=0 0 792 612]{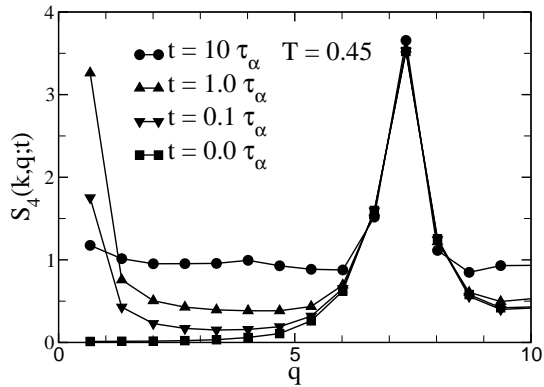}
\caption{\label{s4time}The four-point correlation function $S_4(\vec{k},\vec{q};t)$ 
for $\theta = 0$, where $\theta$ is the angle between $\vec{k}$ and $\vec{q}$,
calculated at $t = 0$, 0.1 $\tau_\alpha$, $\tau_\alpha$ and $10 \tau_\alpha$ for
a temperature $T=0.45$.}
\end{figure}

Functions similar to (\ref{eq:sfactor}) have been used to examine a growing dynamic length
scale in glass forming liquids \cite{Berthier2004,Berthier2006p3,Lacevic2002,Lacevic2003,Toninelli2005}. 
In Fig.~\ref{s4time} we show results similar to those presented in, \textit{e.g.} Ref.~\cite{Lacevic2003}. 
Specifically, we show in $S_4(\vec{k},\vec{q}_{\|};t)$
for $T=0.45$ at times $t = 0$, $0.1 \tau_{\alpha}$, $\tau_\alpha$, and $10 \tau_\alpha$. 
Note that for $t = 0$, $S_4(\vec{k},\vec{q};0) = S(q)$ where
$S(q)$ is the static structure factor for the $A$ particles.
We would like to emphasize that results shown in Fig.~\ref{s4time} are for one specific 
angle between $\vec{k}$ and $\vec{q}$; the angle between $\vec{q}$ and $\vec{k}$ is zero.
It should be noted that for this angle between vectors $\vec{q}$ and $\vec{k}$, 
$S_4$ does not depend on time for $q = k$. This follows from definition (\ref{eq:sfactor});
$S_4(\vec{k},\vec{k};t) = S(k)$ at all times.

The usual interpretation of results shown in Fig.~\ref{s4time} is that the increase of 
$S_4(\vec{k},\vec{q};t)$ at small $q$ values suggests a growing dynamic length scale $\xi(t)$. 
To find the dynamic length
scale, it is common to fit the small $q$ behavior to a functional form and to examine the scaling
of $S_4(\vec{k},\vec{q};t)$ for small $q$. In such a procedure it is implicitly assumed that 
$S_4(\vec{k},\vec{q};t)$ is
isotropic.

\begin{figure}
\includegraphics[width=3.2in,bb=0 0 792 612]{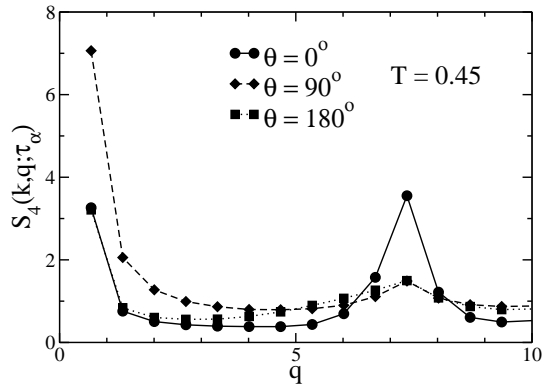}
\caption{\label{sfourangle}The four-point correlation function $S_4(\vec{k},\vec{q};\tau_\alpha)$
for $\theta = 0^\circ$, 90$^\circ$, and 180$^\circ$, where $\theta$ 
is the angle between $\vec{q}$ and $\vec{k}$ calculated for $T=0.45$.}
\end{figure}

However, we find that $S_4(\vec{k},\vec{q};t)$ is not isotropic and depends on the angle 
between $\vec{k}$ and $\vec{q}$.
Shown in Fig.~\ref{sfourangle} is $S_4(\vec{k},\vec{q},\tau_\alpha)$
for $T=0.45$ and for $\theta = 0^\circ$, $90^\circ$, and $180^\circ$ where
$\theta$ is the angle between $\vec{k}$ and 
$\vec{q}$. The anisotropy of $S_4(\vec{k},\vec{q};t)$ adds a complication in finding a unique $\xi(t)$.

Since we do not expect any slowly-decaying with increasing distance 
spatial correlations between self-intermediate scattering functions pertaining to different
particles, we can safely assume that the $\vec{q}\to 0$ limit of $S_4(\vec{k},\vec{q};t)$
is well defined and it does not depend on the angle between vectors $\vec{q}$ and $\vec{k}$.
However, the results shown in Fig.~\ref{sfourangle} suggest that the correlation length 
may be anisotropic. We would like to emphasize that our results are consistent with 
such a possibility but do not prove it. To prove that the correlation length 
is anisotropic one would need to simulate bigger systems in order to be able to examine the
structure factor at smaller wave vectors $\vec{q}$. 

\begin{figure}
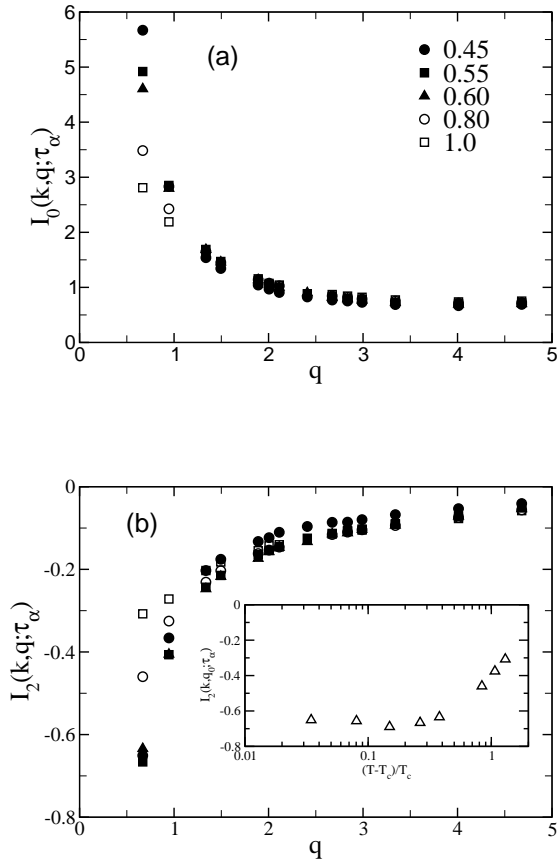

\includegraphics[width=3.2in,bb=0 0 792 612]{I0}
\includegraphics[width=3.2in,bb=0 0 792 612]{I2}
\caption{\label{fig:I0}The projections $I_0(k,q;t)$, (a), and $I_2(k,q;t)$, (b),
as described in the text. The projection
$I_0(k,q;t)$ is the average over angles $\theta$ between $\vec{k}$
and $\vec{q}$ of $S_4(\vec{k},\vec{q};t)$. If $S_4(\vec{k},\vec{q};t)$ does not depend on 
$\theta$, then $I_2(k,q;t)$ would be zero. Shown in the inset is $I_2(k,q_0;\tau_\alpha)$
where $q_0$ is the smallest wave vector allowed due to periodic 
boundary conditions as a function of temperature. The symbols in (a) and (b) correspond
to the same temperatures.}
\end{figure}

We examine the anisotropy of the four-point structure factor by calculating the projection of $S_4(\vec{k},\vec{q};t)$
onto the Legendre polynomials,
\begin{equation}
I_n(k,q;t) = \frac{2n+1}{4 \pi} \int S_4(\vec{k},\vec{q};t) P_n(\hat{\vec{k}} \cdot \hat{\vec{q}})
\mbox{d}\hat{\vec{q}}.
\end{equation}
Shown in Fig.~\ref{fig:I0}(a) is $I_0(k,q;\tau_\alpha)$ (i.e., the angular average of $S_4$)
for $T=1.0$, 0.8, 0.6, 0.55 and 0.45. In most simulational studies of four-point correlation functions
the correlation functions are shown as averages over different directions of wave vector $\vec{q}$,
thus the results are similar
to what is shown in Fig.~\ref{fig:I0}(a). Note, however, that an average over different directions of $\vec{q}$ 
may not correspond to an angular average if the same number of wave vectors corresponding to each angle
between $\vec{q}$ and $\vec{k}$ are not used in the average. Therefore, different routines to 
determine $S_4(\vec{k},\vec{q};t)$ can lead to different conclusions, and our results demonstrate 
that the averaging procedure needs to be performed with caution. 

Shown in Fig.~\ref{fig:I0}(b) is $I_2(k,q;\tau_\alpha)$ for $T=1.0$, 0.8, 0.6, 0.55 and 0.45.
The non-zero values of $I_2$ is a consequence of
$S_4(\vec{k})$ being anisotropic on the $\alpha$ relaxation time scale. The anisotropy is 
largest for the smallest q values. The temperature dependence of $I_2(k,q_0;\tau_\alpha)$
is shown as an inset to Fig.~\ref{fig:I0}(b). The anisotropy 
at the $\alpha$ relaxation time for $q_0$ 
grows with decreasing temperature until around $T=0.5$, then it remains approximately constant.

\subsection{Time dependence of the anisotropy of $S_4(\vec{k},\vec{q}_0;t)$}

\begin{figure}
\includegraphics[width=3.2in,bb=0 0 792 612]{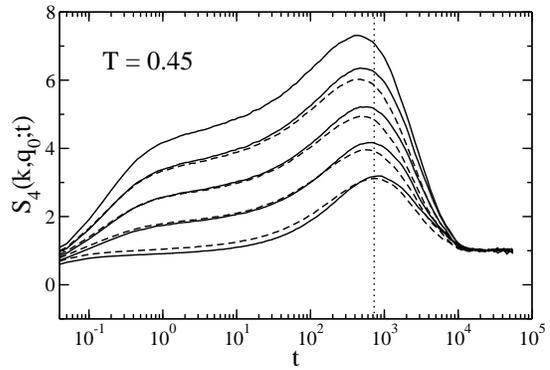}
\caption{\label{fig:fpangle}Time dependence of $S_4(\vec{k},\vec{q}_0;t)$ 
for different angles between $\vec{k}$ and $\vec{q}_0$ calculated for $T=0.45$. The solid lines 
correspond to $\theta = 0^\circ$, 30$^\circ$, 45$^\circ$, 60$^\circ$, and 90$^\circ$
listed from bottom to top. The dashed lines are 120$^\circ$, 135$^\circ$, 150$^\circ$
and 180$^\circ$ listed from top to bottom. The vertical dotted line indicates the $\alpha$ relaxation time. }
\end{figure}

\begin{figure}
\includegraphics[width=3.2in,bb=0 0 792 612]{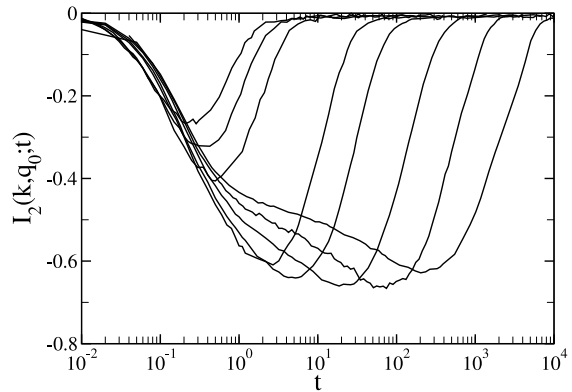}
\caption{\label{fig:ratio}Time dependence of $I_2(k,q_0;t)$ where $q_0$
is the smallest wave vector allowed due to periodic boundary conditions 
for $T=1.0$, 0.9, 0.8, 0.6, 0.55, 0.5, 0.47, and 0.45 listed from left to right.}
\end{figure}

We now turn to the examination of the time dependence of the anisotropy of $S_4(\vec{k},\vec{q};t)$.
To this end, we set $|\vec{q}|$ to be equal to the smallest wave vector allowed for 
our finite size simulation box, $|\vec{q}| = q_0 = 2 \pi/L$, and calculate $S_4(\vec{k},\vec{q}_0;t)$ as a 
function of time for different angles between $\vec{k}$ and $\vec{q}$. Results for $T=0.45$
are shown in Fig.~\ref{fig:fpangle}, and the vertical line marks the $\alpha$ relaxation time. 
We see that $S_4(\vec{k},\vec{q};t)$
grows with increasing time, then reaches a maximum that depends on $\theta$ for a time around the
$\alpha$ relaxation time and finally decays to one at long times.  Note that, while the position of the
maximum is around the $\alpha$ relaxation time, the specific time at which the peak is reached 
depends on the angle between $\vec{k}$ and $\vec{q}$.

To determine the time dependence of the anisotropy, we examined 
$I_2(k,q_0;t)$ where $q_0$ is the smallest wave vector
allowed due to periodic boundary condition, $q_0 = 2 \pi/L$. As seen in Fig.~\ref{fig:ratio}, $I_2(k,q_0;t)$ is zero
at short and long times, but develops a peak at intermediate times. Note
that the shape of $I_2(k,q_0;t)$ is somewhat similar to that of $L_2(k,r_{peak};t)$ 
shown in Fig.~\ref{fig:Ltime} except that $I_2(k,q_0;t)$ is negative (the last fact could be expected
from the relation between $L_2(k,r;t)$ and $I_2(k,q;t)$). The peak height 
increases with decreasing temperature until $T=0.47$, where it starts to decrease. 
However, as we show in the next subsection, 
the correlation length obtained from the fits at $T=0.45$ are all close to 
or greater than half the box length, and it is currently unknown if the
decrease in the peak height is a finite size effect.

\begin{figure}
\includegraphics[width=3.2in,bb=0 0 792 612]{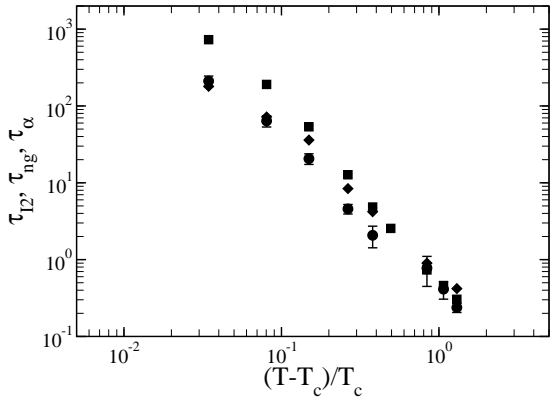}
\caption{\label{fig:ratiopeak}The time corresponding to the maximum value of 
the magnitude of $I_2(k,q_0;t)$, $\tau_{I2}$, 
(circles) compared to the peak position of the non-Gaussian
parameter $\tau_{ng}$ (diamonds) and the $\alpha$ relaxation time $\tau_\alpha$ (squares).}
\end{figure}

To determine when the anisotropy is a maximum at large distances, we 
found the time when $I_2(k,q_0;t)$ reaches its maximum value, $\tau_{I2}$.
Shown in Fig.~\ref{fig:ratiopeak} is the temperature 
dependence of $\tau_{I2}$ (circles) compared to 
$\tau_\alpha$ (squares) and the peak position of the 
standard non-Gaussian parameter $\tau_{ng}$ (diamonds).
We notice similar trends as with the time corresponding to the maximum value of $L_2(k,r_{max};t)$ in that the
$\tau_{I2}$ occurs around $\tau_{ng}$ and has a similar temperature dependence.

\subsection{Effective dynamic correlation length}

There has been some effort to determine the dynamic correlation length by fitting
functions similar to $S_4(\vec{k},\vec{q};t)$ to different functional forms 
\cite{Lacevic2003,Toninelli2005}. Lacevic \textit{et al.}\ \cite{Lacevic2003} used
an Ornstein-Zernicke form $A/(1+(\xi q)^2)$ to fit an overlap function
$S^{ol}_4(q)$ that is isotropic by design,  
while Toninelli \textit{et al.}\ \cite{Toninelli2005} used
$(A-C)/(1+ (\xi q)^\beta) + C$ to fit a function similar to the one studied 
in this work. Lacevic \textit{et al.}\ found a correlation length growing with time 
until the peak time in the associated four-point susceptibility, and then decreasing.
In contrast, Toninelli \textit{et al.}\ found a correlation length growing with time even after the peak in the
associated susceptibility. It is possible that the difference between these findings was related
th the presence of the new parameter $\beta$ in the fit used by Toninelli \textit{et al.}
More recently, Berthier \textit{et al.}\ \cite{Berthier2007p2} used $A/(1+(\xi q)^\beta)$
and found a value of $\beta = 2.4$ provided good fits to the same correlation function studied
in the Ref.~\cite{Berthier2007p2}. Here we focus on a possible anisotropy of the correlation length
at the time equal to the $\alpha$ relaxation time 
and we leave its time dependence for a future study.

We started with
\begin{equation}
S_4(\vec{k},\vec{q};\tau_{\alpha}) = \frac{S_4(\vec{k},0;\tau_{\alpha}) - C}{1 + (\xi_{\theta} q)^2 + (a q)^4} + C,
\label{eq:fit}
\end{equation}
as a fitting function to extract the dynamic correlation length $\xi_{\theta}$.
In Eq.~\eqref{eq:fit} we added a constant $C$ because of the growing baseline which can be seen in
Fig.~\ref{s4time}. We note that since we do not expect any slowly decaying spatial correlations, 
in the limit $q \rightarrow 0$, $S_4(\vec{k},\vec{q};\tau_{\alpha})$ should be 
independent on the angle between $\vec{k}$ and $\vec{q}$. In contrast, in Eq. (\ref{eq:fit}) we
allowed for the dependence of the dynamic correlation length $\xi_{\theta}$ on the angle $\theta$
between $\vec{k}$ and $\vec{q}$.
While fits to Eq.~\ref{eq:fit} were
very good for $q < 3$, the results were not satisfactory. The values of $S_4(\vec{k},0;t)$ were
not consistent for different angles $\theta$ between $\vec{k}$ and $\vec{q}$ and the length 
scales $\xi_{\theta}$ obtained from the fits were greater than 40 at the lowest temperatures. To solve these
problems we performed the procedure described below. We emphasize that simulations of 
larger systems need to be performed to test this procedure and its results.
\begin{figure}
\includegraphics[width=3.2in,bb=0 0 792 612]{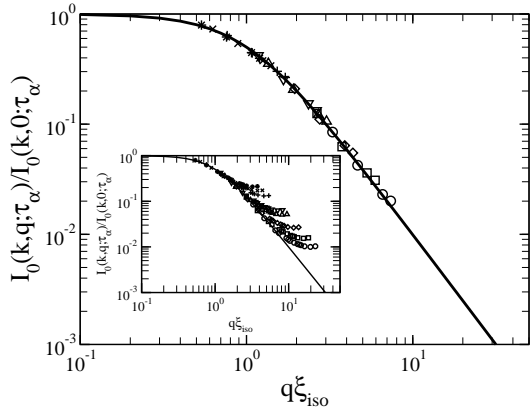}
\caption{\label{fig:I0scale}The four smallest wave vectors 
allowed due to periodic boundary conditions of the projection $I_0(k,q;\tau_\alpha)/I_0(k,0;\tau_\alpha)$
versus $q \xi_{iso}$ for $T=0.45$, 0.47, 0.5, 0.55, 0.6, and 0.8. The solid line is the scaling function $1/(1+x^2)$. 
The inset shows all calculated wave
vectors less than 5.}
\end{figure}

Initially, we attempted to set $C$ and $a$ to zero, thus fitting functions to the Ornstein-Zernicke form. We
set $a$ to zero since it was always very small in the previously attempted fitting procedure. If this 
form is correct, then one could ideally find $S_4(\vec{k},0;t)$ by fitting $S_4(\vec{k},\vec{q};t)$
for different angles between $\vec{k}$ and $\vec{q}$ under the condition that one obtains consistent results. We did not 
obtain consistent results for $S_4(\vec{k},0;t)$ with this procedure and also found that 
we needed to fix the value of 
$S_4(\vec{k},0;t)$ to obtain values of $\xi_{\theta}$ less than 50. Therefore, 
to obtain an estimate for $S_4(\vec{k},0;t)$, we choose to fit $I_0(k,q;t)$ for $q < 1.5$ to 
an Ornstein-Zernicke form and then set the value of $S_4(\vec{k},0;t) = I_0(k,0;t)$ where $I_0(k,0,t)$
is obtained from the fits. Note that this is consistent with our assumption that the limit 
$\lim_{\vec{q}\to 0} S_4(\vec{k},\vec{q};t)$ does not depend on the angle between $\vec{k}$ and $\vec{q}$.
\begin{figure}
\includegraphics[width=3.2in,bb=0 0 792 612]{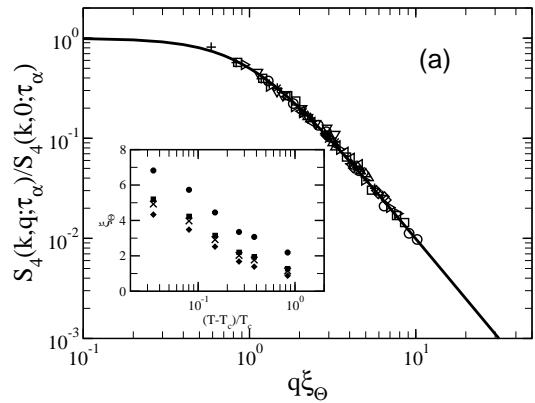}
\includegraphics[width=3.2in,bb=0 0 792 612]{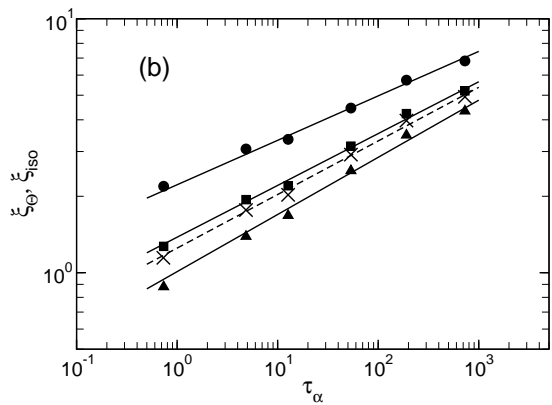}
\caption{\label{fig:s4scale}(a) $S_4(\vec{k},\vec{q};t)/S_4(\vec{k},0;t)$
versus $q \xi_\theta$ for $\theta = 0^\circ$, 45$^\circ$, and 90$^\circ$ where $\theta$ is the angle between $\vec{k}$
and $\vec{q}$ calculated for the temperatures $T=0.8$, 0.6, 0.55, 0.5, 0.47, and 0.45. 
The inset shows the length scales obtained from the different fits
($\theta = 0^\circ$, triangles; $\theta = 45^\circ$, squares; $\theta = 90^\circ$, circles; $\xi_{iso}$, X's).
(b) Dynamic correlation lengths versus the $\alpha$ relaxation time for $\theta = 0^\circ$ (triangles), 45$^\circ$ (squares), 
and 90$^\circ$ (circles). The solid lines are fits to $\xi \sim \tau_\alpha^\gamma$.
The X's and the dashed line corresponds
to $\xi_{iso}$ obtained from the fits of $I_0(k,q;\tau_\alpha)$.}
\end{figure}

If the glass transition is governed by a growing dynamic length scale, then it is expected that
for small enough $\vec{q}$ that $S_4(\vec{k},\vec{q};\tau_{\alpha})/S_4(\vec{k},0;\tau_{\alpha})$ versus $\xi q$ should 
be described by a universal function $F(q \xi)$ that is independent of temperature \cite{Biroli2006}.  
To check if this scaling holds for $I_0(k,q;\tau_{\alpha})$, 
we plotted $I_0(k,q;\tau_{\alpha})/I_0(k,0;\tau_{\alpha})$ versus $q \xi_{iso}$ where
$I_0(k,0;t)$ and $\xi_{iso}$ are obtained from the fits described above and the 
scaling function $1/(1 + x^2)$, which is shown in Fig.~\ref{fig:I0scale}.  
The subscript \textit{iso} in $\xi_{iso}$ emphasizes that this correlation length
was obtained from the orientational average $I_0(k,q;\tau_{\alpha})$ of the four-point structure factor
$S_4(\vec{k},\vec{q};\tau_{\alpha})$.
It appears that this 
scaling holds well for the small $q$ values, but we will again caution that simulations of larger systems 
need to be performed to verify this observation. 
Shown as the inset to the figure is $I_0(k,q;\tau_{\alpha})/I_0(k,0;\tau_{\alpha})$ versus $q \xi_{iso}$ for wave vectors with
a magnitude less than five, and the deviation from the scaling behavior is obvious for the larger 
wave vectors.
The correlation length obtained from $I_0(k,q;\tau_{\alpha})$ 
is on the order of a particle diameter at the larger temperatures, but grows
to about five particle diameters at $T=0.45$. This growth of the
correlation length is consistent with recent results of Berthier and Jack \cite{Berthier2007}.
Note, however, that at the lowest temperature $\xi_{iso}$ is comparable to the half the length of the simulation cell, 
which is the largest length we expect to be able to extract from the simulation without finite size effects.

With the values of $S_4(\vec{k},0;\tau_{\alpha})$ fixed using the
fits from $I_0(\vec{k},0;\tau_{\alpha})$, we fit $S_4(\vec{k},\vec{q};\tau_{\alpha})$ where 
the angle $\theta$ between $\vec{k}$ and $\vec{q}$ are 
0, 45, and 90 degrees to
an Ornstein-Zernicke form where only the correlation length
is allowed to vary. We show $S_4(\vec{k},\vec{q};\tau_{\alpha})/S_4(\vec{k},0;\tau_{\alpha})$ 
versus $q \xi_\theta$, where $\xi_\theta$ depends on the angle $\theta$ between $\vec{k}$
and $\vec{q}$,
for $T = 0.8$, 0.6, 0.55, 0.5, 0.47, and 0.45 in Fig.~\ref{fig:s4scale}. Only wave vectors with
a magnitude less than 1.5 are shown, which corresponds to the four smallest wave vectors
allowed due to periodic boundary conditions
at each temperature and angle. The overlap is
very good for the 18 functions shown, and
shown in the inset to Fig.~\ref{fig:s4scale} 
are the correlation lengths. They depend on the angle between $\vec{k}$ and $\vec{q}$, 
and the correlation lengths are largest for $\theta = 90^\circ$ and smallest for $\theta = 0^\circ$.
Again, we observe that for $\theta = 90^\circ$, the correlation lengths are larger than half the simulation cell
for $T=0.5$ (where $\xi_{90} \approx 4.5$) and lower. This strongly suggests that already at $T=0.5$ simulations of larger 
systems are needed in order to verify the present results. 

In previous studies it has been found that the correlation length is related to the $\alpha$
relaxation time according to a power law, $\xi \sim \tau_\alpha^\gamma$ 
\cite{Whitelam2004,Lacevic2003,Berthier2007p2}. Recently, this behavior was rationalized by
the inhomogeneous mode-coupling theory \cite{Biroli2006}. 
We fitted the the correlation lengths
to a power law of the form $a \tau_\alpha^{\gamma}$ and obtained  
values ranging from $\gamma = 0.22 \pm 0.01$ for $\theta = 0^\circ$ and $\gamma = 0.18 \pm 0.01$ for 
$\theta = 90^\circ$, Fig.~\ref{fig:ratio}. Also shown in Fig.~\ref{fig:ratio}
is $\xi_{iso}$ obtained from $I_0(k,q;t)$; in this case we found 
$\gamma_0 = 0.21 \pm 0.01$, which is very close to the previously reported value of 
0.22, \cite{Whitelam2004}. Using this analysis, we find that the dynamic correlation 
length is not only different for different angles between $\vec{k}$ and $\vec{q}$, but
they also grow at a different rate as the temperature is lowered and
the $\alpha$ relaxation time increases. The range of correlations for particles moving
in the same direction are longer than for particles moving in different directions, but it 
increases slower with decreasing temperature. 

Another scaling prediction is that 
$S_4(\vec{k},0;\tau_\alpha) \sim \tau_\alpha^\Delta$. To test this prediction
we fit $I_0(k,0;\tau_\alpha)$ 
to the form $a \tau_\alpha^\Delta$. In this way we
obtain $\Delta \approx 0.37$, which is again very close to the value of
0.4 reported in Ref.~\cite{Whitelam2004}. These values are slightly smaller than the
recent inhomogeneous mode-coupling theory prediction of $\Delta = 0.5$ \cite{Biroli2006}.

\section{Conclusions}
\label{sec:conclusions}
There have been many studies looking for a growing length scale that accompanies the drastic
slowing down of the dynamics in supercooled and glass forming liquids. Recently, one such possibility was examined by
Biroli \textit{et.\ al} \cite{Biroli2008} where they associated a growing correlation length with a point-to-set
correlation function in a model supercooled liquid. Ever since the
observation of heterogeneous dynamics in supercooled and glassy systems, it has been suggested that a dynamic correlation
length may be associated with the size of the dynamically heterogeneous regions.  
Since two point correlation functions are inadequate 
to describe the correlated motion of atoms and correlated relaxation of the fluid, four-point correlation functions
have been developed to examine this cooperative motion. Normally these correlation functions are assumed
to be isotropic, or are isotropic by design. However, it has been observed that correlated displacements
of particles are not isotropic, and thus it is not surprising that the four-point correlation functions 
might also not be isotropic.

In this work we examined the anisotropy of a four-point correlation function. We found that for distances comparable to 
the nearest neighbor distance the anisotropy initially increases upon supercooling the liquid, but then seems to saturate
or even decrease at the lowest temperatures. Furthermore, the time scale that this anisotropy is a maximum
for nearest neighbor distances is around the $\alpha$ relaxation time at higher temperatures, but then 
it increases slower with decreasing temperature than the $\alpha$ relaxation time and roughly follows
the time corresponding to the peak position of the non-Gaussian parameter $\alpha_2(t)$, $\tau_{ng}$.   

For larger distances, we also found anisotropy of the four-point correlation function. 
We studied the time dependence of this longer ranged anisotropy and found that the time at which it is the
largest also approximately follows $\tau_{ng}$ in the supercooled liquid. 
The longer range anisotropy introduces
a challenge in determining the growing dynamic length scale $\xi$ in glass forming systems. This difficulty is
compounded by the relatively small system sizes usually employed in simulational studies of the glass transition. We developed 
a procedure to extract effective dynamic length scales, but larger system sizes need to be simulated to verify our results.
Our procedure suggests that the dynamic correlation length is different depending on the relative direction
of motion of two particles within the fluid. Furthermore, this anisotropic length scale also increases at a different rate
with decreasing temperature. 

We hope that our present work will stimulate future research in two different directions. First, we advocate the need
to study larger systems and to perform serious finite-size analysis 
of the results \cite{Berthier2003,SteinAndersen2008,SteinPHD}. In particular, 
we expect that in the small $\vec{q}$ limit four-point structure factor $S_4(\vec{k},\vec{q};t)$ is isotropic
and we thus we expect its anisotropic component $I_2(k,q;t)$ to vanish in the small $\vec{q}$ limit. These expectations
should be confirmed by simulations of larger systems. Second, we hope that this work will stimulate a development
of a theoretical model that describes the anisotropy of four-point dynamic correlations.

\section*{Acknowledgments}
We gratefully acknowledge the support of NSF Grant No.~CHE 0517709.

\end{document}